\documentclass[11pt,cls,peerreviewca,onecolumn]{IEEEtran}
\usepackage{hyperref}
\ifCLASSINFOpdf
   \usepackage[pdftex]{graphicx}
\else
\fi
\usepackage[cmex10]{amsmath}
\usepackage{amsfonts,amsthm,amssymb}
\usepackage{mdwmath}
\usepackage{mdwtab}
\usepackage{eqparbox}

\makeatletter
\newcommand*{\rom}[1]{\expandafter\@slowromancap\romannumeral #1@}
\newcommand{\vast}{\bBigg@{4}}
\makeatother

\usepackage{enumerate}
\usepackage{color}
\usepackage{tikz}
\usetikzlibrary{positioning,shapes,calc,shapes.geometric}
\usepackage{pgfplots}
\usetikzlibrary{arrows,intersections}
\usepackage[T1]{fontenc}
\usepackage[utf8]{inputenc}

\IEEEoverridecommandlockouts
\begin{document}
\title{Critical Database Size for Effective Caching}
\author{
Ajaykrishnan N., Navya S. Prem\\
Dept. of Electronics and Communication Engineering\\ 
National Institute of Technology Karnataka\\
\{aja.11ec09, nsp.11ec51\}@nitk.edu.in\\
\and
Vinod M. Prabhakaran, Rahul Vaze\\
School of Technology and Computer Science\\
Tata Institute of Fundamental Research, Mumbai\\
\{vinodmp, vaze\}@tifr.res.in\\}
\maketitle
\begin{abstract}
Replicating or caching popular content in memories distributed across the
network is a technique to reduce peak network loads.  Conventionally, the
performance gain of caching was thought to result from making part of the
requested data available closer to end users.  Recently, it has been shown 
that by using a carefully designed technique to store the contents in the 
cache and coding across data streams a much more significant gain can be 
achieved in reducing the network load. Inner and outer bounds on the network 
load v/s cache memory tradeoff were obtained in~\cite{1}. We give 
an improved outer bound on the network load v/s cache memory tradeoff. 
We address the question of to what extent caching is effective in reducing 
the server load when the number of files becomes large as compared to the number of users. 
We show that the effectiveness of caching become small when the number of files 
becomes comparable to the square of the number of users.
\end{abstract}

\section{Introduction}
In recent times, there has been an increase in demand for online video
streaming leading to high data traffic. Also, it is observed that the demands are 
variable across time, with periods of high and low traffic demand.
The load on the server is high during peak hours when a majority of users access video and relatively 
low at other times.  Thus, there exists the possibility of storing content at
the end users during the off peak hours such that the load on the server is
reduced during peak hours. This method is called {\em caching}. There are two
main phases involved in this process, placement phase and delivery phase. In
the placement phase, data is stored at the end user when the network is
relatively uncongested; here the constraint is the cache memory size at the
user. Also, at this stage the actual request the user might make is not usually
known. In the delivery phase, when the actual requests of the users are made,
the constraint is the rate required to serve all the requested content. 

\par A straightforward approach is to cache a copy of a fraction of all the
files at all the users. Then in the delivery phase, the central server needs to
send only the remaining parts of the requested files.  This is effective
only when the cache size is comparable to the database size at the server.

\par A more sophisticated approach is to allow the central server to
satisfy the request of several users with different demands with a single
multicast stream as was shown in~\cite{1} using the idea of network
coding~\cite{2}. Streams are generated by coding across the different files
requested. This reduces the rate as compared to a conventional caching
scheme. The requested files are decoded from the data stream using the contents
stored in the local cache memory. The gain from this approach is not only
proportional to the cache size but also increases with the increasing number 
of users. Another approach suggested in~\cite{1} is to store contents that are 
coded across files to reduce the rate. 

\par In~\cite{1}, inner and outer bounds on the optimal tradeoff between cache
size $M$ at each user and the data rate $R$ required to service any set of
single file requests from all the users were obtained. Considering a popularity 
distribution on the files, inner and outer bounds on the tradeoff between cache size and
expected load of the shared link was obtained in~\cite{3}. An online version of
this problem was considered in~\cite{4}. In~\cite{5}, a scheme was proposed
where the placement phase is distributed and not centrally controlled by the central server.
In~\cite{6}, a hierarchical system is considered, where caching happens at two
or more levels. 

\par In this paper, we are interested in the case when the database size is
large compared to the number of users. For a fixed cache size, when the
number of files is considerably large compared to the number of users, no
significant gain in the rate can be achieved by any scheme compared to
having no cache. Specifically, we are interested in finding the minimum
number of files beyond which the benefits of caching disappear in the
setting of~\cite{1}. To this end, we first prove a general outer bound on
the optimal $\left(M,R\right)$ tradeoff which generalizes an example
in~\cite{1}. We show that the gains from caching are small
when the number of files is comparable to the square of the number of
users. We then define the pre-constant to the $\Theta{\left(K^2\right)}$ term 
(where $K$ denotes the number of users). Using the improved outer bound we 
obtain a better upper bound to this pre-constant.

\par The rest of the paper is organized as follows. In Section~\rom{2} we
recapitulate the system model proposed in~\cite{1}, and in Section~\rom{3}
we summarize the different caching strategies proposed there.  We derive a
new outer bound on the tradeoff of cache size and rate in Section~\rom{4}
by generalizing an example in~\cite{1}. In Section~\rom{5}, we calculate 
the minimum number of files beyond which benefits of caching become small. 
We finish with a short discussion in Section~\rom{6}.

\section{System Model}
Consider a system (see Fig.~1) with $K$ users connected to the central server
through a shared, error free link. The server has access to the database
containing $N$ files $W_{1},...,W_{N}$, of $F$ bits each, all independent
and uniformly distributed. Each user has access to a cache ${Z}_k$ of size
$MF$ bits for some real number $M$ $\in$ $[0, N]$.
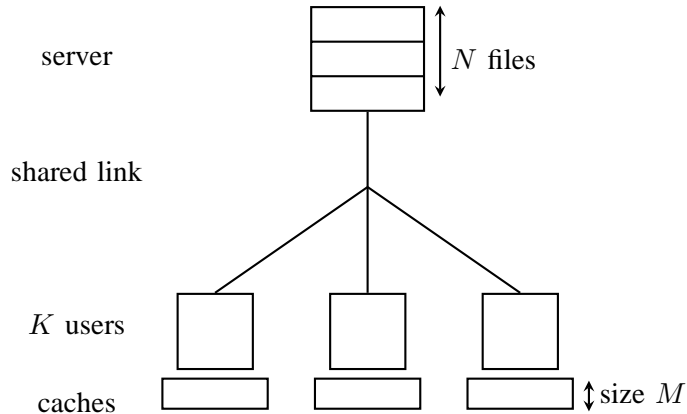
\begin{figure}[h!]
\centering
\tikzset{
block/.style = {rectangle,thick, draw, fill=white!20, 
   text centered, minimum height=3cm,minimum width=1.5cm},
data/.style = {draw,rectangle,thick,minimum height=1cm,fill=white!20,minimum width=1cm},  
user/.style= {draw,thick,minimum height=0.4cm,minimum width=1.4cm},
trans/.style={thick,<->,shorten >=2pt,shorten <=2pt,>=stealth},
}
\begin{tikzpicture}
\node [block,rectangle split, rectangle split parts=3] (nodeA) {};
\node [below = 1cm of nodeA] (spreadout) {}; 
\matrix[
         row sep=2cm,column sep=1cm, below = 1cm of spreadout] (framematrix) {
        \node[data] (frame1){};&
        \node[data] (frame2){};&
        \node[data] (frame3){};\\        
        };
\node [user,below  = .1cm of frame1] (fram1) {};        
\node [user,below = .1cm of frame2] (fram2) {};
\node [user,below = 0.1cm of frame3] (fram3) {};
\draw[thick]  (nodeA) -> (spreadout);   
\node [left =2.5cm of nodeA](name){server};
\node [below =1cm of name](name1){shared link};
\node [below =1.5cm of name1](name2){$K$ users};
\node [below =0.5cm of name2](name3){caches};
\path [draw,thick] (spreadout.north) -- (frame1.north);
\path [draw,thick] (spreadout.north) -- (frame2.north);
\path [draw,thick] (spreadout.north) -- (frame3.north);
 \path[trans] (fram3.north east) ++(0.2,0.03) edge ++(0,-0.55)(fram3.south east)  node[right =0.2cm of nodeA]{$N$ files};
\path[trans] (nodeA.north east) ++(0.2,0.075) edge ++(0,-1.35)(nodeA.south east)  node[right =0.2cm of fram3]{size $M$};
\end{tikzpicture}
\caption{Caching system consisting of $N$ files at the server, $K$ users each having a cache of size $M$ files as in [1].}
\end{figure}
In the placement phase, the user fills the content of its cache by
accessing the database. In the delivery phase, user $k$ requests one of the
files $W_{d_{k}}$ from the database. The server knows all the requests and
transmits a signal $X_{(d_1,...,d_K)}$ of size at most $RF$ bits, where we call $R$
the rate and $(d_1,...,d_K)$ the file request vector. Using the content
${Z}_k$ of its cache and the signal received $X_{(d_1,...,d_K)}$, each user
$k$ must decode its requested file $W_{d_{k}}$. For the rest of the paper
we will be expressing $R$ and $M$ as well as entropies and mutual
informations in units of $F$ bits.

\noindent\textbf{Definition 1}. The memory-rate pair $(M,R)$ is
\textit{achievable} if for every $\varepsilon > 0$ and every large enough
file size $F$ there exists an $(M,R)$ caching scheme such that the 
probability of error in decoding the required file is less than
$\varepsilon$ for each request vector. We define the optimal
\textit{memory-rate tradeoff} as
\[R^*(M)\triangleq \inf \{ R : (M,R) \text{ is achievable}\}.\]

\section{Caching Strategies}
We summarize the three strategies given in~\cite{1}. Here, coding refers to
taking linear combinations of the requested files.

\subsection{Uncoded Caching}\label{subsection:uncoded}
There is no coding involved in this strategy. Each user caches $\frac{M}{N}$
fraction of each file in the placement phase and in the delivery phase the
$1-\frac{M}{N}$ fraction of the file that is not available to the user is transmitted by the server.
Since there are $N$ files, and the size of each file is $F$ bits, the cache
size of each user is $MF$ bits. In the worst case, when no two users request
the same file, for each of the $K$ users, the server needs to transmit the
remaining $1-\frac{M}{N}$ part of each file. This gives an achievable rate $R_U(M)$ which is,
\begin{equation}\label{eq1}
R_U(M)\triangleq K\left(1-\frac{M}{N}\right).\min \left\{1,\frac{N}{K}\right\}.
\end{equation}
There are two factors, $K$ which is the rate without caching and $1-\frac{M}{N}$,
which is the gain because of the availability of caches at the end user
referred to as \textit{local caching gain}. When the number of users is
more than the number of files then an additional gain of $\frac{N}{K}$ is obtained.

\subsection{Coded Caching}\label{subsection:coded}
In this strategy, as mentioned before, the aim is to multicast~(combine various files meant for different users) 
in the delivery phase. In the placement phase, each file is divided 
into $\binom{K}{\frac{MK}{N}}$ equal-sized parts, and each user caches $\frac{MF}{N}$ bits 
of each file such that every $\frac{MK}{N}$ set of users have one part of each file in common.
For the delivery phase, consider any set of $\frac{MK}{N} + 1$ users. Each user in
the set will require a part of the requested file that is present at the 
remaining $\frac{MK}{N}$ users in the set. The central server sends a linear combination
of all the $\frac{MK}{N} + 1$ requested parts. Similar linear combinations are sent
by considering all possible sets of $\frac{MK}{N} + 1$ users. This gives an
achievable rate $R_C{(M)}$ of~\cite{1},
\begin{equation}\label{eq2}
R_{C}(M) \triangleq K.\left(1-\frac{M}{N}\right).\min\left\{\frac{1}{1+\frac{KM}{N}} , \frac{N}{K}\right\}.
\end{equation}
In addition to the \textit{local caching gain} as explained in section
\ref{subsection:uncoded}, coded caching achieves an additional gain of
$\frac{1}{1+\frac{KM}{N}}$ which is the \textit{global caching gain}.
\subsection{Coded Content Placement}
The achievable rate of section~\ref{subsection:coded} can be further
improved by coded content placement. For $\textit{M} = \frac{1}{N}$, coded 
content placement strategy has a lower rate compared to coded caching strategy 
which improves the rate in the region $M=(0,1)$. We illustrate this with an example. 
Consider the case of $N = K = 3$ and $M = 1/3$. In this strategy, we split 
the three files $A,B,C$ into three sub files i.e., $A=(A_1,A_2,A_3)$,
$B=(B_1,B_2,B_3)$ and $C=(C_1,C_2,C_3)$. The caches are stored with $Z_1 =
A_1\oplus B_1\oplus C_1, Z_2 = A_2 \oplus B_2 \oplus C_2$ and $Z_3 = A_3
\oplus B_3 \oplus C_3$.  Consider that user one requests file A, user two
requests file B and user three request file C.  The server satisfies the
requests by transmitting ($B_1,C_1,A_2,C_2,A_3,B_3$) at rate $R = 2$ which
does better than the achievable rate $R_C(M)$ given by~\eqref{eq2} as shown
in Fig.~2.

\section{Lower bound on $R^*(M)$}
In this section, we first summarize the cut-set bound of~\cite{1} and then give an improved bound.
\subsection{Cut-Set Bound}
Let $s\in \{1,...,\min\{N,K\}\}$. Consider $X_{(1,2,\ldots,s)}$, which is
transmitted during the delivery phase, on the shared link when the first $s$ users request files 
$1,2,\ldots,s$, respectively. Then, $X_{(1,2,\ldots,s)}$ along with the caches 
$Z_1,\ldots,Z_s$ of the first $s$ users must determine the files 
$W_1,...,W_s$.  In a similar manner consider $X_{\left(s+1,\ldots,2s\right)},...,
X_{\left(\left(\lfloor{N/s}\rfloor-1\right)s+1,\ldots,\left(\lfloor{N/s}\rfloor s\right)\right)}$. 
Now $X_{\left(1,2,\ldots,s\right)},...,X_{\left(\left(\lfloor{N/s}\rfloor-1\right)s+1,\ldots,\left(\lfloor{N/s}\rfloor
s\right)\right)}$ and $Z_1,....,Z_s$ must determine $W_1,....,W_{\lfloor{N/s}\rfloor s}$. Since 
$\lfloor{N/s}\rfloor$ transmissions of size $R$ and $s$ caches of size $M$ determines 
$s\lfloor{N/s}\rfloor$ files we have,
\begin{equation*}
{\lfloor N/s\rfloor R^*(M) + sM \geq s \lfloor N/s \rfloor}.
\end{equation*}
Solving for $R^*(M)$ and optimizing over all $s$, we obtain
\begin{equation}\label{eq3}
R^*(M) \geq \max \limits_{s \in \{1,.....,\min\{N,K\}\}}\left(s- \frac{s}{\lfloor N/s \rfloor}M\right).
\end{equation}
\subsection{An Improved Bound - An Example}\label{subsection:bound_example}
In this section, we give an example to illustrate how the lower bound
on $R^{*}(M)$ can be tightened compared to the cut-set bound~\eqref{eq3}
by generalizing the approach used in~\cite[Appendix]{1}. 

\noindent\textbf{Example 1.} Consider the case of $N = 9$ files and $K = 4$
users. We consider $X_{1245}, X_{3167}, X_{8912}$ and $X_{7431}$, the
signals transmitted by the server for the request vectors
$(1,2,4,5),(3,1,6,7),(8,9,1,2)$ and $(7,4,3,1)$, respectively. $W_1$ can be
decoded by user $1$ using its cache $Z_1$ and $X_{1245}$. Similarly, user
$2$ can decode file $W_1$ using $Z_2$ and $X_{3167}$.  In the same way,
users $3$ and $4$ can decode file $W_1$ from their caches along with
$X_{8912}$ and $X_{7431}$, respectively. Now, notice that $W_2$ and $W_3$
can be decoded by combining $X_{1245}$, $X_{3167}$ and the caches $Z_1$ of
user $1$ and $Z_2$ of user $2$. Specifically, user $1$ with its cache
$Z_1$ and $X_{3167}$ can decode file $W_3$ and user $2$ with its cache
$Z_2$ and $X_{1245}$ can decode file $W_2$. In the same way, files $W_2$
and $W_3$ can also be decoded by combining $X_{8912}$, $X_{7431}$ and the
caches $Z_3$ of user $3$ and $Z_4$ of user $4$. This combining refers to
step (b) in the chain of inequalities below and is key to obtaining our
lower bound. The remaining files $\left(W_4,W_5,W_6,W_7,W_8,W_9\right)$ can
be decoded by taking all the $4$ request vectors together and using the
corresponding cache of the user that requests that file. The steps given
below demonstrates this procedure. Recall that $R$, $M$, entropies, and
mutual informations are all in units of $F$~bits. For any achievable
memory-rate pair $\left(M,R\right)$, (below we suppress the small
terms resulting from Fano's inequality)
\begin{align*}
4M+ 4R
\geq\; &H(X_{1245},Z_1) + H(X_{3167},Z_2) +
    H(X_{8912},Z_3) + H(X_{7431},Z_4) \\
=\;& H(X_{1245},Z_1|W_1) + I(W_1;X_{1245},Z_1) +
    H(X_{3167},Z_2|W_1) + I(W_1;X_{3167},Z_2) +\\
   & H(X_{8912},Z_3|W_1) + I(W_1;X_{8912},Z_3) +
    H(X_{7431},Z_4|W_1) + I(W_1;X_{7431},Z_4)\\
\stackrel{\text{(a)}}{\geq}\;& H(X_{1245},Z_1|W_1) + H(X_{3167},Z_2|W_1)+
    H(X_{8912},Z_3|W_1) + H(X_{7431},Z_4|W_1) + 4\\
\stackrel{\text{(b)}}{\geq}\;& H(X_{1245},Z_1,X_{3167},Z_2|W_1) +
    H(X_{8912},Z_3,X_{7431},Z_4|W_1) + 4\\
=\;& H(X_{1245},Z_1,X_{3167},Z_2|W_1,W_2,W_3) + 
    I(W_2,W_3;X_{1245},Z_1,X_{3167},Z_2|W_1) +\\
   & H(X_{8912},Z_3,X_{7431},Z_4|W_1,W_2,W_3) +
    I(W_2,W_3;X_{8912},Z_3,X_{7431},Z_4|W_1) + 4\\
\geq\;& H\left(
   \begin{array}{l|l}
   X_{1245},Z_1,X_{3167},Z_2,&W_1,\\X_{8912},Z_3,X_{7431},Z_4 &W_2,W_3 \\
   \end{array}
   \right) +
    I(W_2,W_3;X_{1245},Z_1,X_{3167},Z_2|W_1) +\\
   & I(W_2,W_3;X_{8912},Z_3,X_{7431},Z_4|W_1) + 4 \\
\stackrel{\text{(c)}}{\geq}\;& I\left(
           \begin{array}{r|l}
            W_4,W_5,W_6,W_7,W_8,W_9;&W_1, \\X_{1245},Z_1, X_{3167},Z_2,&W_2,\\ X_{8912},Z_3,X_{7431},Z_4 &W_3
           \end{array}
           \right) + 
           8\\
\stackrel{\text{(d)}}{=}\;& 14,       
\end{align*}  
where (a) follows from Fano's inequality since $W_1$ can be decoded from
each of $(X_{1245},Z_1),(X_{3167},Z_2),\\(X_{8912},Z_3)$ and
$(X_{7431},Z_4)$,
and (b) holds because
\begin{align*}
 H(X_{1245},Z_1|W_1)+ H(X_{3167},Z_2|W_1) &\geq  H(X_{1245},Z_1,X_{3167},Z_2|W_1),\\
H(X_{8912},Z_3|W_1) + H(X_{7431},Z_4|W_1)  &\geq  H(X_{8912},Z_3,X_{7431},Z_4|W_1).
\end{align*}
Similarly (c) follows from Fano's inequality because $W_2,W_3$ can be decoded from each of $(X_{1245},Z_1,
X_{3167},Z_2)$ and $(X_{8912},Z_3,X_{7431},Z_4)$. 
Similarly, (d) holds because $\left(W_4,W_5,W_6,W_7,W_8,W_9\right)$ can be decoded from\\
$(X_{1245},Z_1,X_{3167},Z_2,X_{8912},Z_3,X_{7431},Z_4)$.
Combining the above results we get,
\[M + R^*(M) \geq 3.5.\]
This is an improvement over the cut-set bound which gives $M + R^*(M) \geq 3$.
The coded caching achievable strategy gives $\inf_{M\geq 0} M +
R_C(M)=3.75$ at $M=2.25.$\qed
\subsection{General Lower Bound}
Our main result is the following lower bound on the optimal $(M,R)$
tradeoff. Recall that $M,R$ are in units of $F$~bits.\\ \\
\noindent\textbf{Theorem 1.}\\
For $\alpha > 0$ and $K\geq 2$ users, if $(M,R)$ is achievable,
\begin{enumerate}[(i)]
\item then for $N \geq \left\lceil{\frac{1}{\alpha}}\right\rceil$,
\begin{align}
&\alpha M+R \geq 
\begin{cases}
 \frac{N-\left\lceil{\frac{1}{\alpha}}\right\rceil\left((n-\gamma)^2-(n-\gamma)+1\right)}{2\left\lceil{\frac{1}{\alpha}}\right\rceil(n-\gamma)}+(n-\gamma), &N \leq \left\lceil{\frac{1}{\alpha}}\right\rceil\left(3(n-\gamma)^2 - (n-\gamma) + 1\right)\\
 2(n-\gamma), &N > \left\lceil{\frac{1}{\alpha}}\right\rceil\left(3(n-\gamma)^2 - (n-\gamma) + 1\right)
\end{cases}\label{eq4}
\end{align}
where,
\begin{align}
n &=\vast\lceil{\frac{\left\lceil{\frac{1}{\alpha}}\right\rceil+\sqrt{\left\lceil{\frac{1}{\alpha}}\right\rceil^2+12\left\lceil{\frac{1}{\alpha}}\right\rceil\left(N-\left\lceil{\frac{1}{\alpha}}\right\rceil\right)}}{6\left\lceil{\frac{1}{\alpha}}\right\rceil}}\vast\rceil, \label{eq:n} \\
\gamma &= \max\left(0,\left\lceil n-\frac{K}{2}\right\rceil\right). \label{eq:gamma}
\end{align}
\item then for $N < \left\lceil{\frac{1}{\alpha}}\right\rceil$,
\begin{align}
&\alpha M+R \geq \frac{N}{\left\lceil{\frac{1}{\alpha}}\right\rceil}.
\end{align}
\end{enumerate}
For $\alpha > 1$ and $K\geq 2\left\lfloor\alpha\right\rfloor$ users, if $(M,R)$ is achievable,
\begin{enumerate}[(i)]
\item  then for $N \geq \left\lfloor\alpha\right\rfloor$,
\begin{align}
&\alpha M+R \geq
\begin{cases}
 \frac{N-\left\lfloor\alpha\right\rfloor\left((n-\gamma)^2-(n-\gamma)+1\right)}{2(n-\gamma)}+(n-\gamma)\left\lfloor\alpha\right\rfloor, &N \leq \left\lfloor\alpha\right\rfloor\left(3(n-\gamma)^2 - (n-\gamma) + 1\right)\\
 2(n-\gamma)\left\lfloor\alpha\right\rfloor, &N > \left\lfloor\alpha\right\rfloor\left(3(n-\gamma)^2 - (n-\gamma) + 1\right)
\end{cases}\label{eq4_1}
\end{align}
where,
\begin{align}
 n &=\vast\lceil{\frac{\left\lfloor\alpha\right\rfloor+\sqrt{\left\lfloor\alpha\right\rfloor^2+12\left\lfloor\alpha\right\rfloor\left(N-\left\lfloor\alpha\right\rfloor\right)}}{6\left\lfloor\alpha\right\rfloor}}\vast\rceil, \label{eq:n1} \\
 \gamma &= \max\left(0,\left\lceil n-\frac{K}{2\left\lfloor\alpha\right\rfloor}\right\rceil\right). \label{eq:gamma1}
\end{align}
\item then for $N < \left\lfloor\alpha\right\rfloor$,
\begin{align}
  &\alpha M+R \geq N.
\end{align}
\end{enumerate}
A proof is given in the Appendix. The next example also shows that, 
in general, {Theorem 1} is tighter than the cut-set bound~\eqref{eq3}.

\noindent\textbf{Example 2.}
Consider the case of $N=3$ files and $K=3$ users. The cut-set lower bound~\eqref{eq3}, 
the lower bound of~\eqref{eq4} for $\alpha=1$, and the achievable tradeoffs of~\eqref{eq1} 
and~\eqref{eq2} are shown in Figure~2.
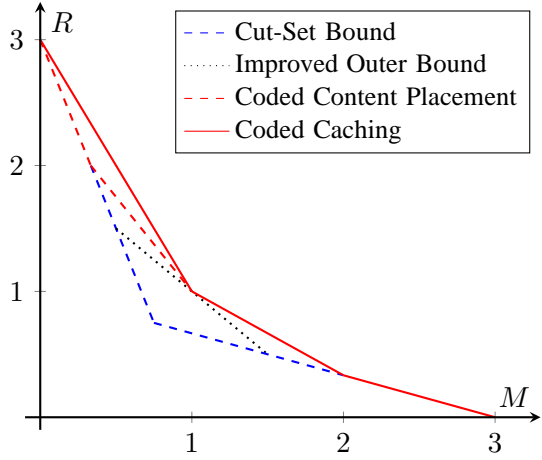
\begin{figure}[h!]
\centering
\begin{tikzpicture}
\begin{axis}[thick,
    legend cell align=left,
    axis x line=middle,
    axis y line=middle,
    axis on top,
    smooth,
    xtick={0,1,2,3},
    xlabel=$M$,
    ylabel=$R$,
    xmin=-0.1, xmax=3.3,
    ymin=-0.1, ymax=3.3,
    legend style={thin,anchor=north east,{font=\small}},
]
\addplot[draw=blue,dashed][domain=(1/3):(3/4)]{3-3*(\x)};
\addplot[draw=black,dotted][domain=0.5:1.5]{2-\x };
\addplot[draw=red,dashed][domain=0:(1/3)]{3-3*(\x)};
\addplot[draw=red][domain=2:3]{1-(1/3)*(\x)};
\addplot[draw=blue,dashed][domain=(3/4):2]{1-(1/3)*\x};
\addplot[draw=red,dashed][domain=(1/3):1]{2.5-1.5*(\x)};
\addplot[draw=red][domain=0.00001:1]{-2*(\x)+3};
\addplot[draw=red][domain=1:2]{(5/3)-(2/3)*(\x)};
\legend{Cut-Set Bound,Improved Outer Bound,Coded Content Placement,Coded Caching}
\end{axis}
\end{tikzpicture}
\caption{The $(M,R)$ tradeoff for $N=3$ files and $K=3$ users. }
\end{figure}
\section{Critical Database Size for Effective Caching}
For any caching system, if the number of files grows we expect the 
reduction in $R$ to be small, for a fixed number of users $K$ and cache size $M$.
In general, each user may find only a small fraction of 
the file requested in its cache.  This results in the server having to send a 
significant part of the requested file in most cases. So the decrease in rate 
$R$ for a fixed $M$ is negligible. Hence, having a large database decreases
the benefits of caching. 

\par To find the minimum database size for a fixed number of users for which 
caching becomes ineffective, we consider the quantity $\left(\alpha M+R^*(M)\right)$, 
which arguably measures the cost of operating a caching system, where
$\alpha > 0$ is the relative cost of cache memory (per user) versus server
bandwidth. Clearly, \[\inf_{M\geq 0} \left(\alpha M + R^*(M)\right) \leq
K,\] since $R^*(M)=K$ for $M = 0$, as the central server must serve the
whole file when there is no cache. We are interested in finding the
smallest size of the database, such that $\inf_{M\geq 0} \left(\alpha M +
R^*(M)\right) = K$.

\noindent\textbf{Definition 2.}
For any $K$ users and $\alpha > 0$, $N(\alpha,K)$ is the minimum number of files such that
\[\inf_{M\geq 0} \left(\alpha M + R^*(M)\right) = K.\]

The following three lemmas give upper and lower bounds on
$N\left(\alpha,K\right)$. {Lemma 1} uses the cut set bound to derive
an upper bound on $N(\alpha,K)$. An improved upper bound using
{Theorem 1} is given in {Lemma 2}. A lower bound on
$N(\alpha,K)$ using the coded caching achievable strategy of~\cite{1} is
given in {Lemma 3}.\\
\noindent\textbf{Lemma 1.}
For $K$ users and $\alpha>0$,\\ 
\[N\left(\alpha,K\right) \leq \left\lceil{\frac{1}{\alpha}}\right\rceil
K^2.\]
\noindent Using the lower bound we derived in Theorem 1, we can improve upon this bound. 
We illustrate this with an example.\\ \\
\noindent\textbf{Example 3.}
Consider the case when there are $K = 4$ users and instead of $9$ files
considered in Section~\ref{subsection:bound_example}, suppose we increase
the number of files to $N=11$.
Following the same procedure as in Example 1, we get
\[M + R^*(M) \geq 4.\]
Thus upper bound on 
\[N(\alpha,K) = 11.\]
This is an improvement compared to $N=16$ files given by {Lemma 1}.
\qed\\ \\
\noindent\textbf{Lemma 2.}
For $K\geq 2$ users and $\alpha>0$,\\
\begin{equation*}
N(\alpha,K) \leq \left\lceil{\frac{1}{\alpha}}\right\rceil \left(3\left\lceil\frac{K}{2}\right\rceil^2 - \left\lceil\frac{K}{2}\right\rceil + 1 \right).
\end{equation*}
For $K\geq 2\left\lfloor\alpha\right\rfloor$ users and $\alpha>1$,\\
\begin{equation*}
N(\alpha,K) \leq \left\lfloor\alpha\right\rfloor \left(3\left\lceil\frac{K}{2\left\lfloor\alpha\right\rfloor}\right\rceil^2 - \left\lceil\frac{K}{2\left\lfloor\alpha\right\rfloor}\right\rceil + 1 \right).
\end{equation*}
\noindent\textbf{Lemma 3.}
For $K$ users and $\alpha>0$,\\
\[N\left(\alpha,K\right) \geq \frac{1}{\alpha}\left(\frac{K^2}{2} +
\frac{K}{2}\right).\]
\noindent The proofs of the lemmas are given in the Appendix. 
From the lemmas it is clear that $N(\alpha,K)=\Theta{\left(K^2\right)}$. 
Thus, it is important to characterize the smallest pre-constant 
to the $\Theta{\left(K^2\right)}$ term which is concretely defined as,
\begin{align*}
\beta_{\alpha} \triangleq \lim_{K \to \infty} \frac{N\left(\alpha,K\right)}{K^2}.
\end{align*}
The following theorem directly follows from the lemmas.

\noindent\textbf{Theorem 2.}
For any $K$ users, $\alpha > 0$ and $N(\alpha,K)$, $\beta_\alpha$ is bounded by\\ 
\begin{align*}
\begin{cases}
    \left(\frac{1}{\alpha}\right) \frac{1}{2} \leq \beta_\alpha \leq \left\lceil{\frac{1}{\alpha}}\right\rceil \frac{3}{4}&,0<\alpha \leq 1\\
    \left(\frac{1}{\alpha}\right) \frac{1}{2} \leq \beta_\alpha \leq \frac{1}{\left\lfloor\alpha\right\rfloor} \frac{3}{4}&,\alpha>1.
\end{cases}
\end{align*}
\par Since the minimum number of files $N(\alpha,K)$ such that $\inf_{M\geq 0}\left(\alpha M + R^*(M)\right)=K$ 
is of $\Theta{\left(K^2\right)}$, we can conclude that the effectiveness of caching becomes small when the 
number of files becomes comparable to the square of the number of users.

\section{Discussion}
In this paper, we consider the case when the number of files is large
compared to the number of users in a caching system.  First, we studied 
inner and outer bounds on the memory-rate tradeoff and present an improved
outer bound by generalizing the approach used in~\cite{1}. We showed that when 
the number of files is comparable to the square of the number of users, the 
benefits of caching become negligible. We defined the $\beta_{\alpha}$ to be the 
pre-constant to the $\Theta{\left(K^2\right)}$ term. Using the improved bound, we 
obtain a better upper bound to this pre-constant.
\par We studied the worst-case shared link load (as in~\cite{1}). We expect 
similar results to hold for the expected load of the shared link under popularity 
distributions on files with a large number of popular files.

\section*{Acknowledgement}
This work was supported in part by Information Technology Research Academy (ITRA), 
Government of India under ITRA-Mobile grant ITRA/15(64)/Mobile/USEAADWN/01.

\appendix
\addcontentsline{toc}{chapter}{Appendix}
\section*{Proof of Theorem 1}
We will first obtain a lower bound on $(M + R)$, for any achievable $(M,R)$, i.e.,
the case of $\alpha=1$. For this, we first consider the case of $K\geq2n$,
where $n$ is as defined in~\eqref{eq:n}. Note that $\gamma$
of~\eqref{eq:gamma} is $0$ in this case. Recall Example 1 where $4$ request
vectors were considered. Similarly, we consider the following $2n$ request
vectors
\begin{subequations}
\begin{align}
     &\left(1,u_1^1,\hdots,u_{n-1}^1,v_1^1,\hdots,v_n^1,t_1^1,\hdots,t_{K-2n}^1\right)\\
     &\left(u_1^2,1,\hdots,u_{n-1}^2,v_1^2,\hdots,v_n^2,t_1^2,\hdots,t_{K-2n}^2\right)\\
     &\vdots \notag\\
     &\left(u_1^n,\hdots,u_{n-1}^n,1,v_1^n,\hdots,v_n^n,t_1^n,\hdots,t_{K-2n}^n\right)\\             
     &\left(v_1^{n+1},\hdots,v_n^{n+1},1,u_1^1,\hdots,u_{n-1}^1,t_1^{n+1},\hdots,t_{K-2n}^{n+1}\right)\\
     &\vdots \notag\\
     &\left(v_1^{2n},\hdots,v_n^{2n},u_1^n,\hdots,u_{n-1}^n,1,t_1^{2n},\hdots,t_{K-2n}^{2n}\right)
\end{align}
\label{eq:request-vectors}
\end{subequations}
Of these, we require that
$1,$ $u_1^1,\ldots,u_{n-1}^1,$ $\ldots,$ $u_1^n,\ldots,u_{n-1}^n$
be distinct. Hence, we will require that $n^2-n+1\leq N$. Furthermore, we
want these along with the $v$'s, i.e., $1,$ $u_1^1,\ldots,u_{n-1}^1,$
 $\ldots,$ $u_1^n,\ldots,u_{n-1}^n,$ $v_1^1,\ldots,v_n^1,$ $\ldots,$
$v_1^{2n},\ldots,v_n^{2n}$ to include all of $1,2,\ldots,N$. 
Hence, we need $n$ to be such that
\begin{align}\label{eq5}
n^2-n+1 \leq N \leq 3n^2 - n + 1.
\end{align}
We can verify that the choice of $n$ in~\eqref{eq:n}, which is reproduced below, satisfies this.
\begin{equation*}
n = \left\lceil\frac{1+\sqrt{1+12(N-1)}}{6}\right\rceil.
\end{equation*}

Consider the first request vector and the first $n$ users. User $1$
requests file $W_1$, and the rest $n-1$ users request files
$\left(W_{u_1^1},\hdots,W_{u_{n-1}^{1}}\right)$.  Similarly, in the second
request vector, user $2$ requests file $W_1$ and the rest $n-1$ users
request files $\left(W_{u_1^2},\hdots,W_{u_{n-1}^{2}}\right)$. In the same
manner for the $n$-th request vector, user $n$ requests file $W_1$ and the
first $n-1$ users request files $\left(W_{u_1^n},\hdots,W_{u_{n-1}^{n}}\right)$. 
These $\left(W_1,W_{u_1^1},\hdots,W_{u_{n-1}^{n}}\right)$ are
$n^2-n+1$ distinct files in the database. For the second set
of $n$ request vectors, users $n+1$ to $2n$ request the same files as users
$1$ to $n$ in the first $n$ request vectors.  For the first $n$ request
vectors, users $n+1$ to $2n$ requests $n^2$ files
$\left(W_{v_1^1},\hdots,W_{v_{n}^{n}}\right)$. For the second $n$ request
vectors, users $1$ to $n$ requests $n^2$ files
$\left(W_{v_1^{n+1}},\hdots,W_{v_{n}^{n+1}}\right)$. By our choices we have
ensured that these $2n^2$ files contain the remaining $N-(n^2 -n+1)$
distinct files.

We now follow the same procedure as in {Example 1}. First file $W_1$ can be
decoded from all the $2n$ request vectors.  This is done by considering the
first request vector and cache $Z_1$, the second request vector and cache
$Z_2$ and so on for the remaining request vectors.  Then, the first set of
$n$ vectors and the second set of $n$ vectors are separately combined to
decode files $\left(W_{u_1^1},\hdots,W_{u_{n-1}^{n}}\right)$.  From the
first $n$ request vectors and caches $\left(Z_1,\hdots,Z_n\right)$ the
files $\left(W_{u_1^1},\hdots,W_{u_{n-1}^{n}}\right)$ can be decoded.
Similarly, from the second set of $n$ vectors and $\left(Z_{n+1},\hdots,
Z_{2n}\right)$ the same set of files can be decoded. The rest $N-(n^2 -n+1)$ 
files which are included in $\left(W_{v_1^1},\hdots,W_{v_{n}^{n}}\right)$ 
can be decoded by considering all the $2n$ request vectors together using 
all the caches $\left(Z_1,\hdots,Z_{2n}\right)$.  All $R$, $M$, entropies 
and mutual informations are in units of $F$ bits and, as before, we suppress 
small terms from Fano's inequality. So for any achievable memory-rate pair 
$(M,R)$ and $K \geq 2n$,
\begin{align*}
2n(M+R) \geq&H\left(X_{\left(1,u_1^1,\hdots,u_{n-1}^1,v_1^1,\hdots,v_n^1,t_1^1,\hdots,t_{K-2n}^1\right)},Z_1\right)+
            \hdots+\\         
            &H\left(X_{\left(u_1^n,\hdots,u_{n-1}^n,1,v_1^n,\hdots,v_n^n,t_1^n,\hdots,t_{K-2n}^n\right)},Z_{n}\right)+\\ 
            &H\left(X_{\left(v_1^{n+1},\hdots,v_n^{n+1},1,u_1^1,\hdots,u_{n-1}^1,t_1^{n+1},\hdots,t_{K-2n}^{n+1}\right)},Z_{n+1}\right)\\
            &+\hdots+H\left(X_{\left(v_1^{2n},\hdots,v_n^{2n},u_1^n,\hdots,u_{n-1}^n,1,t_1^{2n},\hdots,t_{K-2n}^{2n}\right)},Z_{2n}\right)\\                                   
\stackrel{\text{(i)}}{\geq}\;&H\left(
       \begin{array}{l|l}
       X_{\left(1,u_1^1,\hdots,u_{n-1}^1,v_1^1,\hdots,v_n^1,t_1^1,\hdots,t_{K-2n}^1\right)},..,\\X_{\left(u_1^n,\hdots,u_{n-1}^n,1,v_1^n,\hdots,v_n^n,t_1^n,\hdots,t_{K-2n}^n\right)},& W_1 \\Z_1,\hdots,Z_{n}
       \end{array}
       \right)\hspace{0.2cm}+\\                          
       &H\left(
         \begin{array}{l|l}
         X_{\left(v_1^{n+1},\hdots,v_n^{n+1},1,u_1^1,\hdots,u_{n-1}^1,t_1^{n+1},\hdots,t_{K-2n}^{n+1}\right)},\\\hdots,X_{\left(v_1^{2n},\hdots,v_n^{2n},u_1^n,\hdots,u_{n-1}^n,1,t_1^{2n},\hdots,t_{K-2n}^{2n}\right)},&W_1\\
         Z_{n+1},\hdots,Z_{2n}
         \end{array}
         \right)\\
         &+\hspace{0.2cm}2n\\              
\stackrel{\text{(ii)}}{\geq}\; &H\left(
  \begin{array}{l|l}
    X_{\left(1,u_1^1,\hdots,u_{n-1}^1,v_1^1,\hdots,v_n^1,t_1^1,\hdots,t_{K-2n}^1\right)}, \hdots,\\ X_{\left(u_1^n,\hdots,u_{n-1}^n,1,v_1^n,\hdots,v_n^n,t_1^n,\hdots,t_{K-2n}^n\right)} \\
    X_{\left(v_1^{n+1},\hdots,v_n^{n+1},1,u_1^1,\hdots,u_{n-1}^1,t_1^{n+1},\hdots,t_{K-2
    n}^{n+1}\right)}, &{\bf W} \\ \hdots, X_{\left(v_1^{2n},\hdots,v_n^{2n},u_1^n,\hdots,u_{n-1}^1,1,t_1^{2n},\hdots,t_{K-2n}^{2n}\right)} & \\
    Z_1,\hdots,Z_{2n}
  \end{array}
\right)\\
&+\hspace{0.2cm}2n\hspace{0.2cm}+\hspace{0.2cm}2n(n - 1)\\
\stackrel{\text{(iii)}}{\geq}\;&\hspace{0.2cm} 2n^2\hspace{0.2cm}
+\hspace{0.2cm} \left(N-(n^2-n+1)\right),
\end{align*}
where (i) is similar to steps~(a) and~(b) together in Example~1. In step~(ii), which is similar to
step (c) in Example~1. We define
\[{\bf
W}=\left(W_1,W_{u_1^1},\hdots,W_{u_{n-1}^1},\hdots,W_{u_{1}^{n}},\hdots,W_{u_{n-1}^{n}}\right).\]
Step~(iii) is similar to step~(d) of Example~1. Therefore, for $K\geq2n$,
\begin{equation}{\label{eq6}}
 M+R \geq n + \frac{N-(n^2-n+1)}{2n}.
\end{equation}
Notice that $\gamma=0$ for $K\geq2n$, and the definition of $n$ is such that
$N \leq 3n^2 - n + 1$. Thus we have proved the theorem for $\alpha=1$,
$K\geq2n$.

When $K < 2n$, we defined $\gamma \geq 0$ as the smallest integer such 
that $K \geq 2(n-\gamma)$. Notice that since $K\geq2$, $(n-\gamma)>0$. 
Recall that we had considered $2n$ vectors. Now we consider $2(n-\gamma)$ 
request vectors. We follow the same steps as above with $n$ replaced by 
$n-\gamma$. For this, we will now need $N$ to satisfy (cf.~\eqref{eq5})
\[(n-\gamma)^2-(n-\gamma)+1 \leq N \leq 3(n-\gamma)^2 - (n-\gamma) + 1.\]
It is easy to verify that the left inequality follows from the
definitions of $n$ and $\gamma$.
Hence, for $N \leq 3(n-\gamma)^2 - (n-\gamma) + 1$,
\begin{align*}
M+R \geq (n-\gamma)+ \frac{N-((n-\gamma)^2-(n-\gamma)+1)}{2(n-\gamma)}.
\end{align*}
For $K < 2n$ and $N > 3(n-\gamma)^2 - (n-\gamma) + 1$, we proceed as
before, but now the number of files $N$ is larger than the number of
indices $u$'s, $v$'s, and 1. We may set them all to be distinct files and
hence, in step (iii), instead of decoding $N-((n-\gamma)^2-(n-\gamma)+1)$ files,  we now have
$(3(n-\gamma)^2-(n-\gamma)+1)-((n-\gamma)^2-(n-\gamma)+1)$
files. Thus,
\begin{align*}
M+R &\geq  (n-\gamma) +  \frac{(3(n-\gamma)^2-(n-\gamma)+1)-((n-\gamma)^2-(n-\gamma)+1)}{2(n-\gamma)} \\
    &= 2(n-\gamma). 
\end{align*}
This completes the proof for $\alpha=1$.
For generalizing this to any $\alpha>0$, we first consider the case of $N \geq \left\lceil{\frac{1}{\alpha}}\right\rceil$. 
For the case of $K\geq2n$ (i.e., $\gamma=0$), we consider $\left\lceil{\frac{1}{\alpha}}\right\rceil$
sets of $2n$ request vectors similar to~\eqref{eq:request-vectors}. The condition
analogous to~\eqref{eq5} is now
\begin{equation}\label{eq8}
    \left\lceil{\frac{1}{\alpha}}\right\rceil\left(n^2-n+1\right) \leq  N
   \leq \left\lceil{\frac{1}{\alpha}}\right\rceil\left(3n^2-n+1\right),
 \end{equation}
 which can be verified to hold for $n$ as defined in
\eqref{eq:n} with $\gamma=0$.
Now, in step~(i), $\left\lceil{\frac{1}{\alpha}}\right\rceil$ files can be decoded by 
decoding one file from each of $\left\lceil{\frac{1}{\alpha}}\right\rceil$
sets of $2n$ request vectors. Then, in step~(ii), we may now consider
$2\left\lceil{\frac{1}{\alpha}}\right\rceil$ sets of $n$ vectors each
such that $n(n-1)$ files can be decoded from each set. The remaining 
$N-\left\lceil{\frac{1}{\alpha}}\right\rceil\left(n^2-n+1\right)$ can be decoded by combining all the 
$\left\lceil{\frac{1}{\alpha}}\right\rceil2n$ vectors. Hence for $K\geq2n$,
\begin{align*}
2n\left(M+\left\lceil{\frac{1}{\alpha}}\right\rceil R\right)
&\geq \left\lceil{\frac{1}{\alpha}}\right\rceil(2n) + \left\lceil{\frac{1}{\alpha}}\right\rceil\left(2n(n-1)\right) +N-\left\lceil{\frac{1}{\alpha}}\right\rceil(n^2-n+1)\\
&\geq 2\left\lceil{\frac{1}{\alpha}}\right\rceil n^2 +
N-\left\lceil{\frac{1}{\alpha}}\right\rceil(n^2-n+1).
 \end{align*}
Since $\alpha M \geq \frac{M}{\left\lceil{\frac{1}{\alpha}}\right\rceil}$,
we have, for $K\geq2n$,
\begin{equation}\label{eq9}
\alpha M + R \geq
   n + \frac{N-\left\lceil{\frac{1}{\alpha}}\right\rceil(n^2-n+1)}{2\left\lceil{\frac{1}{\alpha}}\right\rceil n}.
\end{equation}
The proof for $K<2n$ is along the same lines as for $\alpha=1$;
as above, we now work with $\left\lceil{\frac{1}{\alpha}}\right\rceil2(n-\gamma)$
request vectors instead of $\left\lceil{\frac{1}{\alpha}}\right\rceil2n$.
\par When $N < \left\lceil{\frac{1}{\alpha}}\right\rceil$ we consider $\left\lceil{\frac{1}{\alpha}}\right\rceil$ request vectors 
such that one of the users, say the first user, requests all $N$ files between these $\left\lceil{\frac{1}{\alpha}}\right\rceil$ request vectors.  
From this we get, $M + \left\lceil{\frac{1}{\alpha}}\right\rceil R \geq N$ which gives $\alpha M+R \geq \frac{N}{\left\lceil{\frac{1}{\alpha}}\right\rceil}$. 
This completes the proof for $\alpha >0$.
\par Now we prove the second part of the Theorem~1 when $\alpha > 1$. We first consider the case of $K\geq2 \left\lfloor\alpha\right\rfloor n$,
where $n$ is as defined in~\eqref{eq:n1}. Note that $\gamma$  
of~\eqref{eq:gamma1} is $0$ in this case. Now consider the following $2n$ request vectors.
\begin{subequations}
\begin{align}
     &\left(1,\hdots,\left\lfloor\alpha\right\rfloor,u_1^1,\hdots,u_{n\left\lfloor\alpha\right\rfloor-\left\lfloor\alpha\right\rfloor}^1,v_1^1,\hdots,v_{n\left\lfloor\alpha\right\rfloor}^1,t_1^1,\hdots,t_{K-2n\left\lfloor\alpha\right\rfloor}^1\right)\\
     &\left(u_1^2,\hdots,u_{\left\lfloor\alpha\right\rfloor}^2,1,\hdots,\left\lfloor\alpha\right\rfloor,u_{\left\lfloor\alpha\right\rfloor+1}^2,\hdots,u_{n\left\lfloor\alpha\right\rfloor-\left\lfloor\alpha\right\rfloor}^2,v_1^2,\hdots,v_{n\left\lfloor\alpha\right\rfloor}^2,t_1^2,\hdots,t_{K-2n\left\lfloor\alpha\right\rfloor}^2\right)\\
     &\vdots \notag\\
     &\left(u_1^n,\hdots,u_{n\left\lfloor\alpha\right\rfloor-\left\lfloor\alpha\right\rfloor}^n,1,\hdots,\left\lfloor\alpha\right\rfloor,v_1^n,\hdots,v_{n\left\lfloor\alpha\right\rfloor}^n,t_1^n,\hdots,t_{K-2n\left\lfloor\alpha\right\rfloor}^n\right)\\             
     &\left(v_1^{n+1},\hdots,v_{n\left\lfloor\alpha\right\rfloor}^{n+1},1,\hdots,\left\lfloor\alpha\right\rfloor,u_1^1,\hdots,u_{n\left\lfloor\alpha\right\rfloor-\left\lfloor\alpha\right\rfloor}^1,t_1^{n+1},\hdots,t_{K-2n\left\lfloor\alpha\right\rfloor}^{n+1}\right)\\
     &\vdots \notag\\
     &\left(v_1^{2n},\hdots,v_{n\left\lfloor\alpha\right\rfloor}^{2n},u_1^n,\hdots,u_{n\left\lfloor\alpha\right\rfloor-\left\lfloor\alpha\right\rfloor}^n,t_1^{2n}\hdots,t_{K-2n\left\lfloor\alpha\right\rfloor}^{2n}\right)
\end{align}
\label{eq:request-vectors2}
\end{subequations}
Of these, we require that $1,\hdots,\left\lfloor\alpha\right\rfloor, u_1^1,\ldots,u_{n\left\lfloor\alpha\right\rfloor-\left\lfloor\alpha\right\rfloor}^1,$ $\ldots,$ $u_1^n,\ldots,u_{n\left\lfloor\alpha\right\rfloor-\left\lfloor\alpha\right\rfloor}^n$
be distinct. Hence, we will require that $\left\lfloor\alpha\right\rfloor(n^2-n+1)\leq N$. Furthermore, we
want these along with the $v$'s, i.e., $1,\hdots,\left\lfloor\alpha\right\rfloor, u_1^1,\ldots,u_{n\left\lfloor\alpha\right\rfloor-\left\lfloor\alpha\right\rfloor}^1,$ $\ldots,$ $u_1^n,\ldots,u_{n\left\lfloor\alpha\right\rfloor-\left\lfloor\alpha\right\rfloor}^n$ 
$v_1^1,\ldots,v_{n\left\lfloor\alpha\right\rfloor}^1,$ $\ldots,$ $v_1^{2n},\ldots,v_{n\left\lfloor\alpha\right\rfloor}^{2n}$ to include all of $1,2,\ldots,N$. 
Hence, we need $n$ to be such that
\begin{align*}
\left\lfloor\alpha\right\rfloor(n^2-n+1) \leq N \leq \left\lfloor\alpha\right\rfloor(3n^2 - n + 1).
\end{align*}
We can verify that the choice of $n$ in~\eqref{eq:n1}, which is reproduced below, satisfies this.
\begin{equation*}
n =\vast\lceil{\frac{\left\lfloor\alpha\right\rfloor+\sqrt{\left\lfloor\alpha\right\rfloor^2+12\left\lfloor\alpha\right\rfloor\left(N-\left\lfloor\alpha\right\rfloor\right)}}{6\left\lfloor\alpha\right\rfloor}}\vast\rceil.
\end{equation*}
Consider the first request vector and the first $n\left\lfloor\alpha\right\rfloor$ users. Users $1$ to $\left\lfloor\alpha\right\rfloor$ 
request files $W_1$ to $W_{\left\lfloor\alpha\right\rfloor}$ and the rest $n\left\lfloor\alpha\right\rfloor-\left\lfloor\alpha\right\rfloor$ 
users request files $\left(W_{u_1^1},\hdots,W_{u_{n\left\lfloor\alpha\right\rfloor-\left\lfloor\alpha\right\rfloor}^{1}}\right)$. Similarly in 
the second request vector, users $\left\lfloor\alpha\right\rfloor + 1$ to $2\left\lfloor\alpha\right\rfloor$ request files $W_1$ to 
$W_{\left\lfloor\alpha\right\rfloor}$ and the rest $n\left\lfloor\alpha\right\rfloor-\left\lfloor\alpha\right\rfloor$ users request 
files $\left(W_{u_1^2},\hdots,W_{u_{n\left\lfloor\alpha\right\rfloor-\left\lfloor\alpha\right\rfloor}^{2}}\right)$. This proceeds in the 
same manner until the $n$-th request vector. These $\left(W_1,\hdots,W_{\left\lfloor\alpha\right\rfloor},W_{u_1^1},\hdots,
W_{u_{n\left\lfloor\alpha\right\rfloor-\left\lfloor\alpha\right\rfloor}^{n}}\right)$ are $\left\lfloor\alpha\right\rfloor(n^2-n+1)$ 
distinct files in the database. For the second set of $n$ request vectors, users $n\left\lfloor\alpha\right\rfloor+1$ to 
$2n\left\lfloor\alpha\right\rfloor$ request the same files as users $1$ to $n\left\lfloor\alpha\right\rfloor$ in the first $n$ request vectors. 
For the first $n$ request vectors, users $n\left\lfloor\alpha\right\rfloor+1$ to $2n\left\lfloor\alpha\right\rfloor$ requests 
$n^2\left\lfloor\alpha\right\rfloor$ files $\left(W_{v_1^1},\hdots,W_{v_{n\left\lfloor\alpha\right\rfloor}^{n}}\right)$. For the second 
$n$ request vectors, users $1$ to $n$ requests $n^2\left\lfloor\alpha\right\rfloor$ files $\left(W_{v_1^{n+1}},\hdots,
W_{v_{n\left\lfloor\alpha\right\rfloor}^{n+1}}\right)$. By our choices we have ensured that these $2n^2\left\lfloor\alpha\right\rfloor$ 
files contain the remaining $N-\left\lfloor\alpha\right\rfloor(n^2 -n+1)$ distinct files. 
\par We now follow the similar procedure as in the case when $0<\alpha\leq 1$. First files $W_1$ to $W_{\left\lfloor\alpha\right\rfloor}$ 
can be decoded from all the $2n$ request vectors. This is done by considering the first request vector and caches $Z_1$ to $Z_{\left\lfloor\alpha\right\rfloor}$, 
the second request vector and caches $Z_{\left\lfloor\alpha\right\rfloor+1}$ to $Z_{2\left\lfloor\alpha\right\rfloor}$ and so on for the 
remaining request vectors. Then, the first set of $n$ vectors and the second set of $n$ vectors are separately combined to decode files 
$\left(W_{u_1^1},\hdots,W_{u_{n\left\lfloor\alpha\right\rfloor-\left\lfloor\alpha\right\rfloor}^{n}}\right)$. From the first $n$ request vectors 
and caches $\left(Z_1,\hdots,Z_{n\left\lfloor\alpha\right\rfloor}\right)$ the files $\left(W_{u_1^1},\hdots,W_{u_{n\left\lfloor\alpha\right\rfloor-\left\lfloor\alpha\right\rfloor}^{n}}\right)$ can be decoded.
Similarly, from the second set of $n$ vectors and $\left(Z_{n\left\lfloor\alpha\right\rfloor+1},\hdots,Z_{2n\left\lfloor\alpha\right\rfloor}\right)$ 
the same set of files can be decoded. The rest $N-\left\lfloor\alpha\right\rfloor(n^2 -n+1)$ files which are included in $\left(W_{v_1^1},\hdots,W_{v_{n\left\lfloor\alpha\right\rfloor}^{2n}}\right)$ 
can be decoded by considering all the $2n$ request vectors together using all the caches $\left(Z_1,\hdots,Z_{2n\left\lfloor\alpha\right\rfloor}\right)$. 
All $R$, $M$, entropies and mutual informations are in units of $F$ bits. So for any achievable memory-rate pair
$(M,R)$ and $K \geq 2n\left\lfloor\alpha\right\rfloor$,
\begin{align*}
2n\left(\left\lfloor\alpha\right\rfloor M+ R\right)
&\geq 2n\left\lfloor\alpha\right\rfloor + 2n(n\left\lfloor\alpha\right\rfloor-\left\lfloor\alpha\right\rfloor) +N-\left\lfloor\alpha\right\rfloor(n^2-n+1)\\
&\geq 2\left\lfloor\alpha\right\rfloor n^2 +
N-\left\lfloor\alpha\right\rfloor(n^2-n+1).
 \end{align*}
 Since $\alpha \geq \left\lfloor\alpha\right\rfloor$, for $K\geq2n\left\lfloor\alpha\right\rfloor$,
\begin{equation*}
\alpha M + R \geq
    n\left\lfloor\alpha\right\rfloor + \frac{N-\left\lfloor\alpha\right\rfloor(n^2-n+1)}{2n}.
\end{equation*}
The proof for $K<2n\left\lfloor\alpha\right\rfloor$ is similar to the case of $0<\alpha\leq1$. 
Here we find the least integer $\gamma$ such that $K\geq2(n-\gamma)\left\lfloor\alpha\right\rfloor$. 
Notice that since $K\geq2\left\lfloor\alpha\right\rfloor$, $(n-\gamma)>0$. Now we consider 
$2(n-\gamma)$ request vectors instead of $2n$. For this, we will now need $N$ to satisfy
\[\left\lfloor\alpha\right\rfloor\left((n-\gamma)^2-(n-\gamma)+1\right) \leq N \leq \left\lfloor\alpha\right\rfloor\left(3(n-\gamma)^2 - (n-\gamma) + 1\right).\]
It is easy to verify that the left inequality follows from the
definitions of $n$ and $\gamma$.
Hence, for \\$N \leq \left\lfloor\alpha\right\rfloor\left(3(n-\gamma)^2 - (n-\gamma) + 1\right)$,
\begin{align*}
\alpha M+R \geq (n-\gamma)\left\lfloor\alpha\right\rfloor+ \frac{N-\left\lfloor\alpha\right\rfloor((n-\gamma)^2-(n-\gamma)+1)}{2(n-\gamma)}.
\end{align*} 
For $K < 2n\left\lfloor\alpha\right\rfloor$ and $N > \left\lfloor\alpha\right\rfloor(3(n-\gamma)^2 - (n-\gamma) + 1)$, we proceed as
before, but now the number of files $N$ is larger than the number of indices $u$'s, $v$'s, and $1,\hdots,\left\lfloor\alpha\right\rfloor$. We may set them all to be distinct files and
hence, in step (iii), instead of decoding $N-\left\lfloor\alpha\right\rfloor((n-\gamma)^2-(n-\gamma)+1)$ files, we now have
$\left\lfloor\alpha\right\rfloor(3(n-\gamma)^2-(n-\gamma)+1)-\left\lfloor\alpha\right\rfloor((n-\gamma)^2-(n-\gamma)+1)$
files. Thus,
\begin{align*}
\alpha M+R \geq  &(n-\gamma)\left\lfloor\alpha\right\rfloor +\frac{\left\lfloor\alpha\right\rfloor(3(n-\gamma)^2-(n-\gamma)+1)-\left\lfloor\alpha\right\rfloor((n-\gamma)^2-(n-\gamma)+1)}{2(n-\gamma)} \\
    = &2(n-\gamma)\left\lfloor\alpha\right\rfloor. 
\end{align*}
This completes the proof for $K<2n\left\lfloor\alpha\right\rfloor$. 
\par When $N < \left\lfloor\alpha\right\rfloor$ we consider $\left\lfloor\alpha\right\rfloor$ caches such that among them all $K$ users are included. 
We consider one request vector where among the users all the $N$ files are requested. Since $N<\left\lfloor\alpha\right\rfloor$ from the $\left\lfloor\alpha\right\rfloor$ caches all the files can be decoded, we get
 $\alpha M + R \geq N.$ 
This completes the proof of Theorem 1 when $\alpha > 1$.

\section*{Proof of Lemmas}
\noindent\textit{Proof of Lemma 1.}\\ \\
Using equation~\eqref{eq3}, by substituting $N = \left\lceil{\frac{1}{\alpha}}\right\rceil K^2$ and $s = K$, 
\begin{align*}
R^*(M) &\geq \left(K - \frac{K}{\lfloor \left\lceil{\frac{1}{\alpha}}\right\rceil K^2/K \rfloor}M\right)\\
\frac{M}{\left\lceil{\frac{1}{\alpha}}\right\rceil} + R^*(M) &\geq K\\
\alpha M + R^*(M) &\geq K
\end{align*}
which gives,
\begin{equation*}
 N(\alpha,K)\leq \left\lceil{\frac{1}{\alpha}}\right\rceil K^2.
\end{equation*}
\qed\\ 
\noindent\textit{Proof of Lemma 2.}\\ \\
This proof follows from {Theorem 1}. Consider the case when $K$ is even and $\alpha>0$. We want to show that for 
\begin{align} \label{eq:lemma2proof}
N=\left\lceil{\frac{1}{\alpha}}\right\rceil\left(\frac{3K^2}{4}-\frac{K}{2}+1\right),
\end{align}
the lower bound of Theorem~1 gives $\alpha M+R \geq K.$
To see this, substitute $N$ from~\eqref{eq:lemma2proof} in~\eqref{eq:n}-\eqref{eq:gamma} to 
see that $n=\frac{K}{2}$ and $\gamma=0$. Then, the lower bound of~\eqref{eq4} indeed gives $\alpha M+R \geq 2n = K$.
Hence we have for even $K$,
\begin{equation*}
 N(\alpha,K) \leq \left\lceil{\frac{1}{\alpha}}\right\rceil\left(\frac{3K^2}{4} - \frac{K}{2} + 1 \right).
\end{equation*}
To handle odd $K$ as well, we note that $N(\alpha,K)$ is a non-decreasing function of $K$ for fixed $\alpha$. Hence for $\alpha>0$ and $K\geq2$,
\begin{equation*}
 N(\alpha,K) \leq \left\lceil{\frac{1}{\alpha}}\right\rceil \left(3\left\lceil\frac{K}{2}\right\rceil^2 - \left\lceil\frac{K}{2}\right\rceil + 1 \right).
\end{equation*}
Following the same procedure for $\alpha>1$ we first consider $K$ to be such that $K=2n\left\lfloor\alpha\right\rfloor$. We choose $N$ to be, 
\begin{equation*}
  N = \left(\frac{3K^2}{4\left\lfloor\alpha\right\rfloor} - \frac{K}{2} + \left\lfloor\alpha\right\rfloor \right).
\end{equation*}
Then, the lower bound of~\eqref{eq4_1} gives $\alpha M+R \geq 2n\left\lfloor\alpha\right\rfloor = K$. 
To find for any $K$, we note that $N(\alpha,K)$ is a non-decreasing function of $K$ for fixed $\alpha$. Hence for $\alpha>1$ and $K\geq2\left\lfloor\alpha\right\rfloor$,
\begin{equation*}
 N(\alpha,K) \leq  \left\lfloor\alpha\right\rfloor \left(3\left\lceil\frac{K}{2\left\lfloor\alpha\right\rfloor}\right\rceil^2 - \left\lceil\frac{K}{2\left\lfloor\alpha\right\rfloor}\right\rceil + 1 \right).
\end{equation*}
Summarizing for $K\geq 2$ users and $\alpha>0$,\\
\begin{equation*}
N(\alpha,K) \leq \left\lceil{\frac{1}{\alpha}}\right\rceil \left(3\left\lceil\frac{K}{2}\right\rceil^2 - \left\lceil\frac{K}{2}\right\rceil + 1 \right).
\end{equation*}
For $K\geq 2\left\lfloor\alpha\right\rfloor$ users and $\alpha>1$,\\
\begin{equation*}
N(\alpha,K) \leq \left\lfloor\alpha\right\rfloor \left(3\left\lceil\frac{K}{2\left\lfloor\alpha\right\rfloor}\right\rceil^2 - \left\lceil\frac{K}{2\left\lfloor\alpha\right\rfloor}\right\rceil + 1 \right).
\end{equation*}

\qed \\
\noindent\textit{Proof of Lemma 3.}\\ \\
To find the minimum number of files such that $\left(\alpha M+R_{C}(M)\right)$ 
is $K$ for the coded caching strategy explained in section~\ref{subsection:coded} notice that,
\begin{align*}
 \alpha M + R_{C}(M) =  &\alpha M + \frac{K\left(1-\frac{M}{N}\right)}{1+\frac{KM}{N}}\\
        =&\alpha M +\frac{KN-KM}{KM+N}.
\end{align*} 
Since $M$ takes only those values for which $\frac{MK}{N} \in \{1,2,\hdots K\}$ as defined by the coded caching strategy we substitute $\frac{MK}{N}=1$. 
Solving this we obtain,
\begin{equation}\label{eq:lemma3}
 N = \frac{1}{\alpha}\left(\frac{K^2}{2} + \frac{K}{2}\right).
\end{equation}
To show that for all $N$ less than~\eqref{eq:lemma3}, the scheme satisfies $\alpha M + R_C(M) < K$, 
consider $N=\left\lceil\frac{1}{\alpha}\left(\frac{K^2}{2} + \frac{K}{2}\right)-1\right\rceil$, 
$M=\frac{N}{K}$ and substitute in $\alpha M+ R_C(M)$. We get, 
\begin{align*}
\alpha M + R_C(M)= &\alpha M + \frac{KN-KM}{KM+N}\\
                    = & \frac{\alpha N}{K} + \frac{K(1-1/K)}{2}\\
                    < & \frac{\alpha(K^2+K)}{2\alpha K} + \frac{K-1}{2}\\
                    < &\frac{K+1}{2} + \frac{K-1}{2}\\
                    < &K.
\end{align*}
Hence,\\
\[N(\alpha,K) \geq \frac{1}{\alpha}\left(\frac{K^2}{2} + \frac{K}{2}\right).\]
\qed
\end{document}